\documentclass[twocolumn,showpacs,preprintnumbers,amsmath,amssymb]{revtex4}

\usepackage{graphicx}% Include figure files
\usepackage{dcolumn}% Align table columns on decimal point
\usepackage{bm}% bold math
\usepackage{color}

\begin{document}

\preprint{YITP-09-27}

\title{Ultrahigh-Energy Photons as a Probe of Nearby Transient Ultrahigh-Energy Cosmic-Ray Sources and Possible Lorentz-Invariance Violation}
% Force line breaks with \\

\author{Kohta Murase}
%\email{kmurase@yukawa.kyoto-u.ac.jp}%
\affiliation{%
Yukawa Institute for Theoretical Physics, Kyoto University, Kyoto 606-8502, Japan
}

\date{July 29}% It is always \today, today,
               
\begin{abstract}
Detecting neutrinos and photons is crucial to identifying the sources of
ultrahigh-energy cosmic rays (UHECRs), especially for transient sources.
We focus on ultrahigh-energy $\gamma$-ray emission from transient
sources such as $\gamma$-ray bursts, since $> \rm EeV$ $\gamma$ rays
can be more direct evidence of UHECRs than $\sim \rm PeV$ neutrinos and
GeV-TeV $\gamma$ rays. 
We demonstrate that coincident detections of $\sim 1-100$ events can be
expected by current and future UHECR detectors such as Auger and 
JEM-EUSO, and the detection probability can be higher than
that of neutrinos for nearby transient sources at $\lesssim 50-100$ Mpc.
They may be useful for constraining the uncertain cosmic radio
background as well as knowing the source properties and maximum energy
of UHECRs. 
They can also give us more than ${10}^{4}$ times stronger limits
on the Lorentz-invariance violation than current constraints.  
\end{abstract}

\maketitle
The origin of ultrahigh-energy cosmic rays (UHECRs) is
one of the biggest mysteries in astroparticle physics, 
and a number of scenarios have been theoretically proposed so far 
(for reviews, see, e.g., \cite{BS00}). 
However, physical conditions in these potential sources are 
uncertain, and observational progress in source identification 
has been limited by the scarcity of experimental data 
(e.g., \cite{NW00}).
The recent results of large area detectors such as 
the Pierre Auger Southern Observatory (PAO) have started to give us 
crucial clues to the origin. 
Indeed, the first PAO results reported a significant 
correlation between the arrival directions of the highest-energy 
cosmic rays and the large-scale structure of the Universe, which is 
inhomogeneous up to dozens of Mpc (e.g., \cite{PAO07,KW08}).  
However, not only active galactic nuclei (AGNs) \cite{RB93,AD08} but also 
transient sources such as $\gamma$-ray bursts (GRBs) \cite{Wax95,Mur+06} 
and magnetars \cite{Aro03} can be UHECR sources so far.
Even if the association of UHECRs with AGNs is real, the PAO report 
suggests that the majority of the correlating AGNs seems
radio-quiet, a class of objects not 
showing any nonthermal high-energy emission, 
and the power of those AGNs seems insufficient to produce UHECRs \cite{Geo+08}. 
This problem may be solved if UHECRs are produced during active states such as flares \cite{FG09,AD01}. 
When the UHECR sources are transient, the magnetic fields in the Universe not only deflect UHECRs but also 
cause significant time delays compared to photons and neutrinos
generated during the bursts (e.g., \cite{MT09}). 
Then, due to difficulties in identifying the sources through UHECRs, 
it is more favorable to detect photons and neutrinos. 
 
We focus on ultrahigh-energy (UHE) photon emission from 
transient UHECR sources with numerical calculations 
considering the cosmic infrared, microwave, or radio background (CIB/CMB/CRB)
and the loss due to the intergalactic magnetic field (IGMF). 
We demonstrate that UHE photons can be the most useful messenger
for nearby sources, 
though the results depend on source properties and the uncertain CRB. 
Constraints on the Lorentz-invariance violation (LIV) are also discussed.

\textit{UHE photon production in the source.---} 
If cosmic rays are accelerated up to ultrahigh energies,  
hadronic $\gamma$ rays and neutrinos should be produced via the $pp$ or 
$p \gamma$ reactions, but their efficiency and resulting spectra 
depend on source models \cite{Mur+06,AD01,WB97,Mur07}. 
In this work, for demonstrative purposes, we mainly consider 
$p \gamma$ photons and neutrinos from GRBs \cite{Mes06} as an example 
(e.g., \cite{Mur+06,WB97,Mur07}).
We especially demonstrate the case of relatively low luminous bursts, 
motivated by recent suggestions that nearby bursts such as  
GRB 060218 are dimmer but more numerous than classical GRBs [and they are often called 
low-luminosity (LL) GRBs] \cite{Cam+06}.   
Other cases such as AGN flares can also be considered similarly.

First, we write a source UHECR energy spectrum as 
\begin{equation}
\tilde{\mathcal E}_{\rm CR}^{\rm iso} 
\equiv E_p^2 \frac{d N_{p}^{\rm iso}}{d E_p} \approx 
\frac{1}{\rho} 
{\left( E_p^2 \frac{d \dot{N}_{\rm CR}}{d E_p} \right)}_{E_p=E_0} 
{\left( \frac{E_p}{E_{0}} \right)}^{2-p} {\rm e}^{-E_p/E_p^{\rm max}},
\label{UHECR} 
\end{equation}
where $\rho$ is the local apparent rate of bursts responsible for the 
observed UHECRs, $p$ is the source spectral index, and 
$E_p^{\rm max}$ is the maximum UHECR energy. 
In this work, assuming proton composition, 
we adopt $p=2$ expected in the ankle scenario \cite{BS00,NW00}.
The energy input rate at $E_0={10}^{19}$ eV is estimated as 
$E_p^2 \frac{d \dot{N}_{\rm CR}}{d E_p} \sim 
{10}^{44}~{\rm erg} {\rm Mpc}^{-3} {\rm yr}^{-1}$ from 
the UHECR data \cite{BS00,Wax95,MT09}. 
The recent PAO results suggest that, if 
the UHECR sources are transient, the UHECR energy input per burst at ${10}^{19}$ eV 
is $\tilde{\mathcal E}_{\rm HECR}^{\rm iso} \equiv 
\tilde{\mathcal E}_{\rm CR}^{\rm iso} ({10}^{19}~{\rm eV})
\sim {10}^{50.5}~{\rm erg}~\rho_{2.5}^{-1}$ 
($0.1~{\rm Gpc}^{-3} {\rm yr}^{-1} \lesssim \rho 
\lesssim {10}^{3.5}~{\rm Gpc}^{-3} {\rm yr}^{-1}$) \cite{MT09}. 
Classical GRBs correspond to $\rho \sim 0.1-1
~{\rm Gpc}^{-3} {\rm yr}^{-1}$ while LL GRBs, hypernovae 
\cite{Cam+06} and AGN flares \cite{FG09} may correspond to 
$\rho \sim {10}^{2-3}~{\rm Gpc}^{-3} {\rm yr}^{-1}$. 

Provided a proton spectrum and a target photon spectrum, 
we can calculate spectra of $p \gamma$ photons and neutrinos.
As a photon spectrum, we use a (broken) power law 
which is also expected in the synchrotron emission mechanism: 
$d n/d\varepsilon \propto \varepsilon^{-\alpha}$. 
Here $\varepsilon$ is 
the target photon energy in the comoving frame 
(while $\varepsilon_{\rm ob} \approx \Gamma \varepsilon$ 
is the energy in the observer frame, where 
$\Gamma$ is the bulk Lorentz factor). 
In the case of GRB prompt emission, $\alpha \sim 1$ 
for $\varepsilon < \varepsilon^b$ 
and $\alpha \sim 2$ 
for $\varepsilon^b < \varepsilon$ 
are observed as typical values, 
where $\varepsilon^b$ is the break energy \cite{Mes06}. 
Then, using the $\Delta$-resonance approximation,
the effective optical depth for the $p \gamma$ reaction 
in the source is estimated as \cite{Mur+06,WB97} 
$f_{p \gamma} \approx t_{\rm dyn}/t_{p \gamma}
\sim 0.1
\frac{L_{\gamma,48}^b}{r_{14.5} \Gamma_{1.5}^2 \varepsilon_{\rm ob, 10~keV}^b}
{(E_p/E_p^b)}^{\alpha-1}$,
where $E_p^b \approx 1.6 \times {10}^{16}~{\rm eV}~\Gamma_{1.5}^2 
{(\varepsilon_{\rm ob, 10~keV}^b)}^{-1}$
is the resonance energy, $L_{\gamma}^b$ is the photon luminosity at $\varepsilon_{\rm ob}^b$, 
$r$ is the emission radius, $t_{\rm dyn} \approx r/\Gamma c$ is the 
dynamical time scale of the relativistic source, and $t_{p \gamma}$ is the 
$p \gamma$ energy loss time scale.
We may expect efficient meson production 
(${\rm min}[1, f_{p \gamma}] \sim 0.01-1$). 

The produced pions decay into $\gamma$ rays and neutrinos via
$\pi^0 \rightarrow 2 \gamma$ and $\pi^{\pm} \rightarrow 
e^{\pm}+{\nu}_{e}({\bar{\nu}}_{e})+{\nu}_{\mu}+{\bar{\nu}}_{\mu}$.
Lifetimes of $\pi^0$ and $\pi^{\pm}$ are 
$8.4 \times {10}^{-17}$ and $2.6 \times {10}^{-8}$ s, respectively.
Because of $\pi^0$'s very short lifetime, we may expect that 
sufficiently high-energy $\gamma$ rays reflect proton and photon 
spectra, leading to $E_\gamma^2 \phi_{\gamma}^{\rm pri} 
\propto f_{p \gamma} E_{\gamma}^{2-p} \propto 
E_{\gamma}^{1+\alpha-p}$. 
However, for $\gamma$ rays, it is an important issue 
whether they can escape from the source without 
significant source attenuation.
The most relevant process is pair creation, 
whose optical depth is evaluated for the same photon 
field as that given for the $p \gamma$ reaction \cite{Mur+06,WB97} 
(see also Ref. \cite{Gra+08} for more general discussions). 
But, for synchrotron sources, the self-absorption becomes 
important at low energies \cite{RMZ04,LW07}. 
In the case of GRB prompt emission, the synchrotron 
self-absorption energy 
is roughly estimated as $\varepsilon_{\rm ob}^{sa} \sim 
2~{\rm eV}~{(L_{\gamma,48}^b)}^{1/3} L_{M,49}^{1/3} \xi_{B}^{1/3} 
\Gamma_{1.5}^{-2/3} r_{14.5}^{-1}$, where 
$L_{M}$ is the outflow luminosity and $\xi_B$ is the ratio of 
the magnetic energy density to the photon energy density \cite{Mes06,RMZ04}.  
When the Klein-Nishina effect is relevant above 
$\tilde{\varepsilon}_{\rm ob}^{sa}$ (where 
$\tilde{\varepsilon}_{\rm ob} \equiv \Gamma^2 m_e^2 c^4/\varepsilon_{\rm ob}$),
we have \cite{Mur+06,WB97,RMZ04,LW07}
\begin{eqnarray}
\tau_{\gamma \gamma} 
\simeq 50 \left( \frac{f_{p \gamma}(E_p^b)}{0.1} \right)
\left\{ \begin{array}{ll}
{\left( \frac{E_{\gamma}}{\tilde{\varepsilon}_{\rm ob}^{b}} \right)}^{\alpha-1} 
& \mbox{($E_{\gamma} \leq \tilde{\varepsilon}_{\rm ob}^{sa}$)} \\
{\left( \frac{\tilde{\varepsilon}_{\rm ob}^{sa}}{\tilde{\varepsilon}_{\rm ob}^{b}} \right)}^{\alpha-1}
{\left( \frac{E_{\gamma}}{\tilde{\varepsilon}_{\rm ob}^{sa}} \right)}^{-1}
\Lambda 
& \mbox{($\tilde{\varepsilon}_{\rm ob}^{sa} < E_{\gamma}$)}
\end{array} \right. 
\label{gg}
\end{eqnarray}
where $\Lambda$ is the logarithmically energy-dependent term from 
the Klein-Nishina effect. An example for a somewhat bright LL GRB-like burst 
is shown in Fig. 1, where accurate
cross sections of $\gamma \gamma \rightarrow 
e^{+} e^{-}$ and $\gamma e^{-} \rightarrow e^{-} e^{+} e^{-}$ are
used. Although the escapability depends on source models, 
UHE photons could escape from the source at 
$E_{\gamma}^{\rm thin} \sim {10}^{16}~{\rm eV}~L_{\gamma,48}^b
r_{14.5}^{-1} {(\varepsilon_{\rm ob, 10~keV}^b 
\varepsilon_{\rm ob, 1~eV}^{sa})}^{-1} 
{(\tilde{\varepsilon}_{\rm ob}^{sa}/\tilde{\varepsilon}_{\rm ob}^{b})}
^{\alpha-1} \Lambda$ unless additional 
low-energy photon fields exist.
In this work, we calculate primary $\gamma$-ray spectra by exploiting elaborate
numerical calculations including various processes \cite{Mur+06,Mur07}
and the result for the somewhat bright LL GRB-like burst is shown in Fig. 2, where  
$\xi_B=1$ and the other relevant parameters  
are described in the caption of Fig. 1. 
In the calculations, we also estimate the maximum energy and 
$E_{p}^{\rm max} \simeq {10}^{20.5}$~eV is obtained in this case. 
Roughly speaking, the primary $\gamma$-ray spectrum can be approximated as 
$E_{\gamma}^2 \phi_{\gamma}^{\rm pri} \approx \frac{1}{4 \pi D^2} \frac{1}{2}
f_{p \gamma} e^{-(E_{\gamma}^{\rm thin}/E_\gamma)} 
E_p^2 \frac{d N_p^{\rm iso}}{d E_p}$, 
where the typical $\gamma$-ray energy is $E_{\gamma} \approx 0.1 E_p$.

Next, we briefly discuss the case of AGN flares. Following Ref. \cite{FG09},
let us adopt $L_{\gamma}^b={10}^{45}~{\rm erg}~{\rm s}^{-1}$, 
$r={10}^{16.5}$~cm and $\Gamma={10}^{0.5}$ 
(corresponding to the duration of $T \sim
{10}^{4-5}$~s). We can obtain $E_{\gamma}^{\rm thin} \sim
{10}^{16.5}~{\rm eV}
{(\tilde{\varepsilon}_{\rm ob}^{sa}/\tilde{\varepsilon}_{\rm ob}^{b})}
^{\alpha-1} \Lambda$, taking $\varepsilon_{\rm ob}^b \sim 10$~eV and 
$L_{M} \sim {10}^{47}~{\rm erg}~{\rm s}^{-1}$.
Then, escape of UHE photons from the source is possible 
but it typically seems more difficult than the case of GRBs. 
The meson production efficiency is also estimated as 
$f_{p \gamma} \sim 0.1 {(E_p/E_p^b)}^{\alpha-1}$, so
that the expected fluence level of primary UHE photons can be similar to that
shown in Fig. 2.   
Although detailed results depend on scenarios and parameters, 
the relevant processes are similar, and it is enough to
show the case of GRBs for the demonstrative purpose of this work. 
 
In order to prove acceleration of UHECRs, 
detections of particles with $E_{\nu} \approx 5~{\rm EeV}~E_{p,20}$ or 
$E_{\gamma} \approx 10~{\rm EeV}~E_{p,20}$ are favorable. 
However, very high-energy neutrino emission may be suppressed 
since charged mesons and muons can cool down before they decay
\cite{WB97}. It will be true especially in the case of GRB prompt
emission since the comparison between $\pi^{\pm}$'s lifetime and 
its synchrotron cooling time gives 
$E_{\nu}^{\rm syn} \approx 4.7 \times {10}^{17}~{\rm eV}
~{\xi}_B^{-1/2} {(L_{\gamma,48}^{b})}^{-1/2} \Gamma_{1.5}^2 r_{14.5}$ 
above which the flux is suppressed by 
${(E_{\nu}/E_{\nu}^{\rm syn})}^{-2}$.

\begin{figure}[bt]
\includegraphics[width=0.8\linewidth]{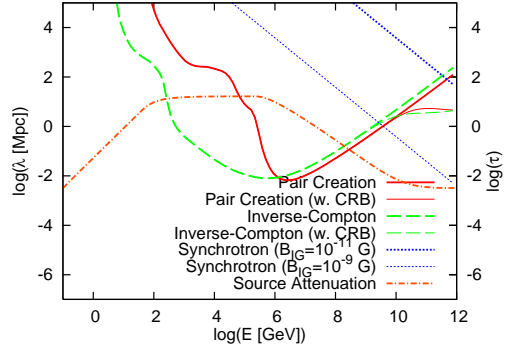}
\caption{\small{\label{Fig1}
The interaction and attenuation lengths of high-energy photons and 
electron-positron pairs propagating in the Universe.   
An example of the source optical depth of photons in the case of 
bright LL GRB-like bursts is also shown for demonstration, 
where $r={10}^{15}$ cm, $\Gamma={10}^{1.5}$, $L_{\gamma}^b=
{10}^{48}~{\rm erg}~{\rm s}^{-1}$, $\alpha=1$~and~$2.2$, 
$\varepsilon_{\rm ob}^b=10$ keV and $\varepsilon_{\rm ob}^{sa}=10^{0.5}$ eV. 
}}
\end{figure}

\textit{Processes outside the source.---} 
UHE photons, even if they can escape from the source, cannot avoid 
attenuation by the CIB, CMB, and CRB \cite{BS00}.
At the high energies of $\gtrsim {10}~{\rm PeV}$, the attenuation
lengths for pair creation and inverse-Compton scattering in the CMB are  
roughly $\lambda_{\gamma \gamma} \sim 2~{\rm Mpc}~E_{\gamma,18}/{\rm
ln}(400 E_{\gamma,18})$ and $\lambda_{\rm IC} \sim 2~{\rm
Mpc}~\gamma_{e,12}/[{\rm ln}(1800 \gamma_{e,12})-2]$, respectively, 
and numerically calculated lengths are shown in Fig. 1. 
The CRB has rather large uncertainty 
at present, so that we consider the extreme two cases: the non-CRB case and 
the case of the high CRB model developed in Ref.  \cite{PB96}. 
Secondary electron-positron pairs generated by pair creation are still 
energetic and upscatter cosmic background photons. 
These boosted photons can create pairs as long as they are energetic, 
and the process repeats itself until the energy of degraded photons is in
the 1-10 TeV range. Hence, as a result of this cascade process, 
the effective attenuation lengths are longer than 
the original ones \cite{BS00}.  
To take into account this cascade effect, 
we have solved cascade equations \cite{BS00,AD08}, whose results 
agree with previous works \cite{Pro86,MAN07}. 
We can neglect double pair creation and Bethe-Heitler
processes when $E_{\gamma}^{\rm max} \lesssim {10}^{21}$ eV \cite{BS00}.

In Fig. 2, the resulting UHE $\gamma$-ray spectra are demonstrated 
for the numerically calculated primary $\gamma$-ray spectrum. 
Cascaded $\gamma$ rays with $\lesssim {10}^{19.5}$ eV can enhance our 
chance to detect UHE signals from nearby transient sources (see below). 
For $D \sim 40~{\rm Mpc}$, the $\gamma$-ray fluence is 
$E_{\gamma}^2 \phi_{\gamma} \sim 
{10}^{-6.5}~{\rm erg}~{\rm cm}^{-2}~f_{p \gamma,-1}
\tilde{\mathcal E}_{\rm HECR, 50.5}^{\rm iso}$ at $\sim {10}^{19.5}$ eV in 
the non-CRB case, allowing us to expect their detections 
if a UHECR burst occurs at $\sim 3$ Mpc (like Cen A), 
at $\sim 20$ Mpc (like the Virgo cluster),
and at $\sim 40$ Mpc (like GRB 980425).
For $D \sim 20~\rm Mpc$, we have $\mathcal{N} 
\sim 10~{\rm events}~f_{p \gamma,-1} 
\tilde{\mathcal E}_{\rm HECR, 50.5}^{\rm iso}$ by PAO  ($A \sim 3000~{\rm km}^2$).  
But, the results depend on 
the uncertain CRB, which could make detections 
difficult for bursts at $\gtrsim 50$ Mpc. 
They are also affected by the maximum UHECR energy. 
\begin{figure}[tb]
\includegraphics[width=0.8\linewidth]{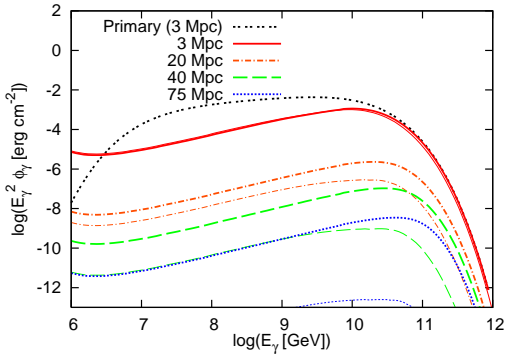}
\caption{\small{\label{Fig2}
Energy fluences of UHE photons from a LL GRB-like UHECR burst 
with $\tilde{\mathcal{E}}_{\rm HECR}^{\rm iso}
={10}^{50.5}$ erg 
for each distance. 
The primary $\gamma$-ray spectrum is also shown (see the caption of
Fig. 1 for the source parameters).
Thick lines show the non-CRB case while thin lines show 
the CRB case, with $B_{\rm IG}={10}^{-13}$ G. 
The burst rates expected within each distance are 
$1/28000$, $1/94$, $1/12$, and $1/1.8~{\rm yr}^{-1}$.
}}
\end{figure}

The number of events $\mathcal{N}$ would not usually be large, 
so that space and time coincidence with low-energy photons 
(e.g., x/$\gamma$ rays) is important.
Since the magnetic deflection angle is $\theta_B \approx \lambda_{\rm
IC}^{1/2} \lambda_{\rm coh}^{1/2}/r_L \sim 2.6 \times {10}^{-6} B_{\rm
IG, -13} \lambda_{\rm coh, \rm kpc}^{1/2} \gamma_{e,13}^{-1/2}/
{[{\rm ln}(18000 \gamma_{e,13})-2]}^{1/2}$, the magnetic time delay, which 
is typically the most important, is ${\Delta t}_B \approx 
\frac{1}{4} \frac{D}{c} \theta_B^2 \sim 860~{\rm s}~D_{40 \rm Mpc} 
B_{\rm IG,-13}^2 \lambda_{\rm coh, \rm kpc} \gamma_{e,13}^{-1}$ 
\cite{BS00,MAN07}. 
Hence, as long as the IGMF is weak enough, the magnetic time delay can
be shorter than the burst duration of $T$ (e.g., $\sim {10}^{2-3}$~s
for GRBs), and coincident detections of cascaded UHE photons can be expected.   
Note that such weak IGMFs are possible in voids, and 
the mean free path of UHE photons is $\gtrsim$~a few Mpc so that 
UHE photons may escape from the structured region (filaments and clusters)
and UHE pairs may feel weak IGMFs only \cite{Mur+06,MAN07}.
On the other hand, UHECRs can have longer and sufficient time delays 
since they should feel stronger IGMFs in the structured region 
($\sim {\rm nG}-\mu{\rm G}$) and the galactic magnetic field \cite{MT09,Pla95}.
If IGMFs are not weak or if there is the possible magnetic field
of $\sim 0.1~\mu \rm G$ in the galactic halo, we expect coincidence
only for noncascaded photons, and cascaded photons (especially for
$\lesssim {10}^{19.5}$~{\rm eV} photons) spread the signals 
out in time. The energy dependence is critical here, and lower-energy GeV-TeV photons have the much longer duration \cite{MAN07}. 
\begin{figure}[tb]
\includegraphics[width=0.8\linewidth]{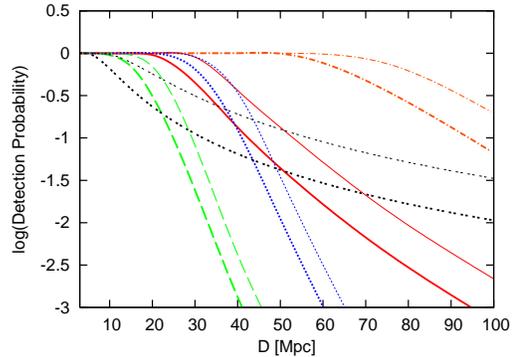}
\caption{\small{\label{Fig3} 
The comparison of Poisson probabilities to detect UHE ($> 10~\rm EeV$) 
photons and high-energy ($> 10~\rm PeV$) neutrinos from a LL GRB-like UHECR burst. 
For UHE photons, 
$A=3000~{\rm km}^2$ without the CRB (solid lines),
$A=3000~{\rm km}^2$ with the CRB (dashed lines),
$A=3 \times {10}^5~{\rm km}^2$ without the CRB (dotted-dashed lines),
and $A=3 \times {10}^5~{\rm km}^2$ with the CRB (dotted lines).
For neutrinos, 
$A=1~{\rm km}^2$ (double-dashed lines), assuming IceCube-like detectors. 
Thick and thin lines are for $\tilde{\mathcal E}_{\rm HECR}^{\rm iso}
={10}^{50.5}~{\rm erg}$ and $\tilde{\mathcal E}_{\rm HECR}^{\rm iso}
={10}^{51}~{\rm erg}$, respectively.
}}
\end{figure}

In Fig. 3, we compare the Poisson probability 
($\mathcal{P}=\Sigma_n \mathcal{N}^n {\rm e}^{-\mathcal{N}}/n!$) 
to detect $\geq 1$ events  
for neutrinos by ${\rm km}^{3}$ telescopes such as IceCube 
with that for UHE photons by large area detectors such 
as PAO and JEM-EUSO ($A \sim$ a few $\times {10}^5~{\rm km}^2 $) \cite{Ahr+04}. 
Spectra of both neutrinos and UHE photons are calculated for the same
source parameters used in Figs. 1 and 2.  
UHE photons can be more useful 
to prove transient UHECR sources at from 
$\sim 10$ Mpc to $\sim 50-100$ Mpc.  

The burst rate of transient UHECR sources 
within 100 Mpc is estimated from $\rho$  
as $\sim 1.3~{(\tilde{\mathcal{E}}_{\rm HECR, 50.5}^{\rm iso})}^{-1}~{\rm yr}^{-1}$
\cite{MT09}.  
In fact, LL GRBs, hypernovae, and AGN flares may have 
corresponding rates of $\rho \sim {10}^{2-3}~{\rm Gpc}^{-3} {\rm yr}^{-1}$ \cite{Mur+06,FG09}. 
The expected rate is not so high, but there is still room
to detect signals in the future.  

\textit{Implications and discussions.---}
In this work, we have demonstrated that,
for nearby sources within dozens of Mpc, 
detections of UHE photons by PAO and JEM-EUSO 
can be expected and are important to identify the transient UHECR
sources. They can also be useful to test the LIV which is often expected in
quantum gravity theories \cite{Cam+98}. 
Let us expand the energy-dependent light velocity as 
$c^{\prime} = c {(E_{\gamma}/\zeta_n E_{\rm pl})}^n$, where $E_{\rm pl}$ is the Planck energy.
Then, the LIV-induced time delay is written as 
${\Delta t}_{\rm LIV} \simeq (D/c){(E_{\gamma}/\zeta_n E_{\rm
pl})}^{n}$.
When UHE photons are coincident with low-energy photons during $T$, 
from ${\Delta t}_{\rm LIV} < T$, we obtain bounds of
$\zeta_{1} \gtrsim 3.4 \times {10}^{3} E_{\gamma,19} T_{3}^{-1} D_{40 \rm Mpc}$ for $n=1$ 
and
$\zeta_{2} \gtrsim 1.7 \times {10}^{-3} E_{\gamma,19}  T_{3}^{-1/2} D_{40 \rm Mpc}^{1/2}$ for $n=2$. The current limits by Fermi observations of GRB 080916C are $
\zeta_1 \gtrsim 0.13$ and $\zeta_2 \gtrsim 7.9 \times {10}^{-10}$
\cite{Fer09}. Hence, possible detections of 
UHE photons may give us the most stringent limits on the LIV, 
as well as possible EeV neutrinos \cite{LW07}. 
Even when observed UHE photons have time delays, we could potentially
constrain the LIV since the energy dependence of $\Delta t_{\rm LIV}$
is different from that of $\Delta t_B$. 
Furthermore, LIV-induced modifications to
the attenuation may increase our chance to detect UHE photons 
\cite{Cam+98}.

UHE photons have two merits compared to neutrinos, 
in that (1) $\gtrsim$ EeV neutrinos may be suppressed due to the meson cooling
and their detections via Earth-skimming $\nu_{\tau}$'s 
may not be so easy \cite{Ahr+04}, and (2) $\sim$ PeV neutrinos suitable for 
IceCube-like detectors directly suggest acceleration of $\sim 100$ PeV cosmic rays rather 
than UHECRs ($\gtrsim {10}^{18.5}$~eV). 
On the other hand, UHE $\gamma$-ray 
fluences depend on the CRB 
and the spectral shape at the highest energies. 
Another uncertainty comes from $f_{p \gamma}$ 
and $\tau_{\gamma \gamma}$ (e.g., in the mixed-composition scenario, 
it would be more difficult to detect neutrinos and photons \cite{Mur+06}).  
Conversely, detections of UHE photons will give us important 
information, e.g., enabling us to constrain the CRB. 

The background UHE photons from the sources would not be so important 
compared to cosmogenic photons 
and current PAO limits \cite{PAO07}, though the future anisotropy search
via, e.g., finding multiplet events could be relevant 
especially if the duration of UHE photon emission is long. Their 
arrival distribution may also be expected to trace the matter 
distribution of the nearby Universe. 

Finally, let us discuss associated GeV-TeV $\gamma$-ray signals. 
They should also be important since a 
significant fraction of UHE photons 
should be radiated as lower-energy $\gamma$ rays via 
the synchrotron or inverse-Compton emission.  
But results would be rather sensitive to the IGMF in voids.
If the IGMF is so weak, they may be detected as 
pair echoes, i.e., long lasting cascaded $\gamma$-ray 
emission \cite{MAN07}.  
If the IGMF is not weak, they may be 
detected as a pair halo \cite{ACV94} but their flux 
should be greatly reduced.  
If the IGMF is strong enough, UHE pairs will 
emit $\sim {\rm GeV}~\gamma_{e,13}^2 B_{\rm IG, -9}$
photons that could be detected by Fermi, though 
they are also be contaminated by accompanied leptonic components.

K.M. thanks C.D. Dermer, H. Takami, U. Jacob, T. Nakamura, S. Inoue, 
V.S. Berezinsky, and T. Ebisuzaki and especially appreciates comments by K. Ioka. 
He acknowledges support by a Grant-in-Aid from JSPS and by a Grant-in-Aid for 
the Global COE Program ``The Next Generation of Physics, Spun from Universality and  Emergence" from MEXT.

\appendix

\newpage %Just because of unusual number of tables stacked at end

\end{document}